\documentclass[a4paper]{aa}

\usepackage{psfig,times}
\begin{document}

\newcommand{\bra}[1]{\langle #1\rangle}
\newcommand{\chk}[1]{[{\em check: #1}]}

\def\ni{\noindent} 
\def\ea{{et\thinspace al.}}                        
\def\eg{{e.g.}\ }                                    
\def\ie{{i.e.}\ }                                   
\def\cf{{cf.}\ } 
\def\rot{\mathop{\rm rot}\nolimits}
\def\div{\mathop{\rm div}\nolimits} 
 
\newcommand{\block}{\vrule width 0.5 true cm height 6pt depth 0pt \ }
\def\lgl{\langle}
\def\rgl{\rangle}
\def\nbl{{\bf \nabla}}
\def\om{\omega}
\renewcommand{\vec}[1]{\mbox{\boldmath$#1$}}
\def\blll{\par\vskip 36pt\noindent}
\def\alf{$\alpha$}
\def\L{$\Lambda$}
\def\Om{\Omega}
\def\nT{$\nu_T$\ }
\def\mT{$\mu_T$\ }
\def\cT{$\chi_T$\ }

\def\og{\hbox{\,$g$}\kern-0.43em\raise.32ex\hbox{$^\circ$}}
\def\vog{\hbox{\,$\vec{g}$}\kern-0.43em\raise.32ex\hbox{$^\circ$}}
\def\ogf{\hbox{\,$g$}\kern-0.43em\raise.32ex\hbox{$^\circ$}_f}
\def\ogi{\hbox{\,$g$}\kern-0.43em\raise.32ex\hbox{$^\circ$}_i}
\def\ogm{\hbox{\,$g$}\kern-0.43em\raise.32ex\hbox{$^\circ$}_m}
\def\ogj{\hbox{\,$g$}\kern-0.43em\raise.32ex\hbox{$^\circ$}_j}
\def\ogomi{\hbox{\,$\Omega$}\kern-0.53em\raise.52ex\hbox{$^\circ$}_i}
\def\ogom{\hbox{\,$\Omega$}\kern-0.53em\raise.52ex\hbox{$^\circ$}_j}
\def\om{\hbox{\,$\Omega$}\kern-0.53em\raise.52ex\hbox{$^\circ$}}

\def\apj{{ApJ}\ }       
\def\apjs{{ Ap. J. Suppl.}\ } 
\def\apjl{{ Ap. J. Letters}\ } 
\def\pasp{{ Pub. A.S.P.}\ } 
\def\mn{{MNRAS}\ } 
\def\aa{{A\&A}\ } 
\def\aasup{{ Astr. Ap. Suppl.}\ } 
\def\baas{{ Bull. A.A.S.}\ } 
\def\csss{{Cool Stars, Stellar Systems, and the Sun}\ }
\def\an{{Astron. Nachr.}\ }
\def\sp{{Solar Phys.}\ }   
\def\gafd{{Geophys. Astrophys. Fluid Dyn.}\ } 
\def\ass{{Ap\&SS}\ }
\def\acta{{Acta Astron.}\ }
\def\jfm{{J. Fluid Mech.}\ }
\def\K{Kitchatinov}
\def\GR{G\"unther R\"udiger\ }
\def\AIP{Astrophysikalisches Institut Potsdam}
\def\gR{G. R\"udiger}
\def\R{R\"udiger}
\def\B{Brandenburg}
\def\abs{\par\bigskip\noindent}            
\def\qq{\qquad\qquad}                      
\def\qqq{\qquad\qquad\qquad}               
\def\q{\qquad}
\def\F{Ferri\`{e}re}
\def\DR{differential rotation\ }
\def\bib{\item{}}
\def\top{\item}
\def\toptop{\itemitem}
\def\start{\begin{itemize}}
\def\stop{\end{itemize}}
\def\beg{\begin{equation}}
\def\ende{\end{equation}}



\newcommand{\BS}{$\backslash$}
\let\oldverbatim\verbatim
\renewcommand{\verbatim}{\expandafter\small\oldverbatim}


\title{Lithium as a passive tracer  probing the rotating solar 
tachocline turbulence} 

\author{G. R\"udiger\inst{1,3} \and  V.V. Pipin\inst{1,2}} 
\offprints{G. \R}
\institute{Astrophysikalisches Institut Potsdam, An der
Sternwarte 16, D-14482 Potsdam, Germany
\and
Institute for Solar-Terrestrial Physics,
P.O. Box 4026, Irkutsk 664033, Russia
\and
Department of Mathematics, University of Newcastle upon Tyne,
NE1 7RU, UK}
\date{\today}
\markboth{G. R\"udiger \& V.V. Pipin: Lithium
probing    the tachocline turbulence}
{G. R\"udiger \& V.V. Pipin: Lithium   probing    the 
tachocline turbulence}

\abstract{
The rotational influence on the eddy-diffusivity  tensor $D_{ij}$ for
anisotropic turbulence fields is considered in order to explain
the lithium decay law during the spin-down process of
solar-type stars.  Rotation proves to be highly effective in
the transfer of chemicals through the solar tachocline (beneath the
convection zone) which is assumed to contain only turbulence
with horizontal motions. The effect is so strong that the
tachocline  turbulence must not exceed a limit of
$\sim\!\!10^{-(3\dots4)}$ of the rms velocity in the convection
zone in order to let the lithium survive after 
Gigayears. Such long depletion times can also be explained by
a very small rotational  influence upon the eddy-diffusion
tensor if it is realized with correlation times shorter
than 15 min. It is argued that such slow and/or
short-living turbulence  beneath the convection zone could hardly
drive the solar dynamo. \\
In our theory the  diffusion  remains small for rapid rotation due to the 
rotational quenching of the turbulence. In young stellar clusters like 
Pleiades, there should be a (positive) correlation between rotation rate and 
lithium abundance, where the  fastest stars should have maximal lithium. First inspections of the data seem to confirm this finding.}    
\maketitle


\section{Introduction}
\begin{figure}
\psfig{figure=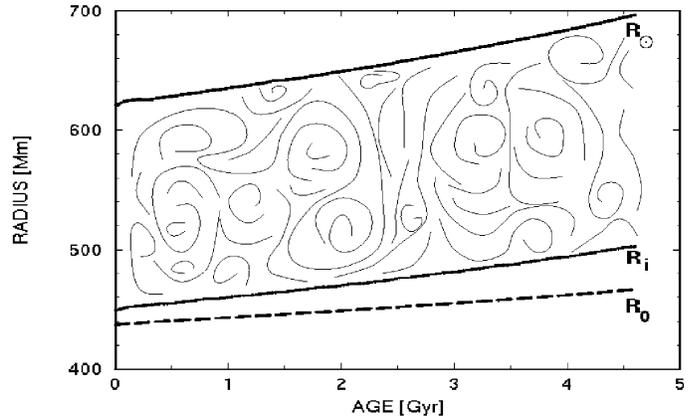,height=5.5truecm,width=9truecm,angle=270}
\caption
{The outer structure of the Sun during its main-sequence life.
$R_i$ limits the unstably stratified outer convection zone
while $R_0$ displays the lithium burning zone with a temperature
of 2.6 Mio K. Computed by  
S. K\"ammerer (2000) with  the solar model of Stix \& Skaley (1990)} 
\label{tacho2}
\end{figure}

According to Duncan (1981), the  lithium at the
surface of cool main-sequence stars decays exponentially, unlike the stellar 
spin-down, which follows a  power law
$t^{-1/2}$ (Skumanich 1972). The characteristic decay time is
about 1 Gyr. The primordial lithium is destroyed by
nuclear reactions  at temperatures in excess   of $2.6\!\cdot\!
10^6$ K (see Michaud \& Charbonneau 1991; Ahrens \ea\ 1992), \ie about 40.000 km below the lower edge of the
convection zone (Fig. \ref{tacho2}).  There must be a  drift process for  the
chemicals from the bottom of the convection zone through the
solar  `tachocline' to the burning domain. The effect must  be
small, however, in order  to allow the existence of
lithium in the solar atmosphere even after 4.6 Gyr.  The
lithium decay time is about $10^7$ times the  convection zone
diffusion time of $\sim\!\! 100$ yrs.  We are thus looking for a
rather small effect which, however, cannot simply be
microscopic diffusion (Schatzman \& Maeder 1981; Spruit 1984; Zahn 1989, 1992). 
In the following, the
consequences are presented of a  quasilinear mean-field
approximation of an anisotropic turbulence field which might be
located in the solar `tachocline', which is stably stratified
unlike the unstably-stratified solar convection zone. However, 
to form reduction
factors of $10^{7...8}$ for any eddy diffusivity is a
nontrivial problem. In other words, the desired eddy
diffusion coefficient only exceeds the molecular one by two
orders of magnitude (Baglin \ea\ 1985; Lebreton \& Maeder 1987)
and this is not easy to explain (Vincent \ea\ 1996).

The transport of a  passive scalar  is governed by the diffusion equation,
\beg
{\partial \rho C \over \partial t} + {\rm div} \left(\rho C \vec{u} - \rho
D \vec{\nabla} C\right) = 0,
\label{2}
\ende
with $D$ as the microscopic diffusion coefficient. In the sense
of the anelastic approximation we shall always apply the
source-free condition of the mass flux, i.e. 
\beg
{\rm div} \rho \vec{u} = 0.
\label{3}
\ende
If the field $\vec{u}$ describes a turbulent flow pattern, the fields are 
split into a   mean and a fluctuating part, 
\beg
C = \bar{C} + C', \q \vec{u} = \bar{\vec{u}} + \vec{u}' .
\label{4}
\ende
Averaging (\ref{2}) we get the well-known diffusion equation in
the presence of turbulence
\beg
{\partial \rho \bar{C} \over \partial t} + {\rm div} \left\{\rho
\langle C'\vec{u}'\rangle - \rho D \vec{\nabla}
\bar{C}\right\} = 0 .
\label{5}
\ende
The influence of a meridional circulation considered by
Charbonneau \& Michaud (1991), Zahn (1992) and
Chaboyer \& Zahn (1992) is neglected here. 

The
fluctuations of the chemical concentration in a quasilinear
approximation follow the relation 
\beg
{\partial \rho C' \over \partial t} + {\rm div}\left(\rho \bar{C}\vec{u}'
- \rho D \vec{\nabla} C'\right) = 0.
\label{6}
\ende
From this we have to compute the turbulent concentration
flux vector $\langle C'\vec{u}'\rangle $ which must be inserted
into (\ref{5}) in order to find the mean-field diffusion
equation. In particular, we are interested in the effect of the
basic stellar rotation and its 
consequences during the stellar spin-down process (see Charbonnel \ea\ 1992). 
\section{The diffusion tensor}
In  a corotating
frame of reference the equation for the fluctuating part of the velocity field  $ \vec u' $ is 
\beg
{\partial \vec{u}' \over \partial t} + 2\vec{\Omega} \times
\vec{u}'  + {1\over \rho}  \nabla p'  
 - \nu_{\rm t} \Delta \vec{u}' 
= \vec{f}',
\label{8}
\ende
with $\vec f'$ as the random turbulence force and $\nu_{\rm t}$ as
some background eddy viscosity. The desired correlation between
concentration fluctuations and velocity fluctuations, $\langle
C'\vec{u}'\rangle$,  
can be found by Fourier 
transforming (\ref{8}) using
\beg
\vec{u}'(\vec{x},t) = \int \hat{\vec{u}}(\vec{k},\omega)
e^{{\rm i}({\bf k}\cdot {\bf x} - \omega t)} d\vec{k}\, d\omega . 
\label{9}
\ende
As usual, the original anisotropic turbulence is assumed to be stationary,
homogeneous and anisotropic, i.e.
\begin{eqnarray}
\lefteqn{\langle \hat u_j^{(0)}(\vec{k},\omega)
\hat u_j^{(0)}(\vec{k}',\omega')\rangle =  \delta(\omega +
\omega') \delta(\vec{k} + \vec{k}')}\nonumber\\
&& \bigg\{{E(k,\omega) \over 16\pi k^2} \left(\delta_{ij} - k_i^\circ
k_j^\circ\right) + 
  {3\over 16\pi k^2} E_2(k,\omega)\bigg(\delta_{ij}
- k_i^\circ k_j^\circ -  \nonumber\\
&&  -(\vec{g}\cdot \vec{k}^\circ)^2
\delta_{ij} - 
 g_i g_j + (\vec{g}\cdot \vec{k}^\circ) (g_i k^\circ_j + g_j
k_i^\circ)\bigg)\bigg\} 
\label{10}
\end{eqnarray}
(cf. R\"udiger 1989). $E(k,\omega)$ is
the isotropic part of the turbulence spectrum, $\vec{k}^\circ \equiv
\vec{k}/k$ and $\vec{g}$ is the vertical unit vector. The radial
turbulence intensity may be denoted by $w^2= \langle
u_r^{(0)2}\rangle$ while for the azimuthal turbulence intensity 
$v^2= \langle u_\phi^{(0)2}\rangle$.  
An anisotropy parameter $s$ is defined  by
$v^2 = s w^2$, so that a large $s$ denotes horizontal-type turbulences.  It
is 
\beg
w^2 = {1\over 3}\int\!\!\!\int\limits_{\!\!\!0}^{\!\!\!\infty} E(k,\omega) dk
d\omega .
\label{11}
\ende
The vertical vector $\vec g$  represents the basic anisotropy, which
is described by the spectrum $E_2(k,\omega)$ 
of {\em additive} horizontal motions
\beg
 v^2 = {1\over 2} \ \int\!\!\!\int\limits_{\!\!\!0}^{\!\!\!\infty}
E_2(k,\omega) dk d\omega.
\label{12}
\ende
The result for  the turbulent concentration-flux vector may be
written as an anisotropic diffusion in terms of the mean concentration
gradient, i.e.
\beg
\langle C' u_i'\rangle = - D_{ij} \ {\partial \bar{C} \over
\partial x_j} 
\label{16}
\ende
(cf. Dolginov \& Silantev 1992). We shall compute in  the following the diffusion tensor
without rotation and with rotation for the isotropic and anisotropic
parts of the  turbulence fields. 

\subsection{Isotropic turbulence}
The result is very simple without rotation. It follows
\beg
D_{ij} = {1\over 3}  \int\!\!\!\int\limits_{\!\!\!0}^{\!\!\!\infty} {D k^2 \over
\omega^2 + D^2 k^4}\  E(k,\omega) \ dk d\omega\ \delta_{ij},
\label{171}
\ende
or just
\beg
D_{ij} =  D_{\rm T} \ \delta_{ij} 
\label{172}
\ende
with the eddy diffusivity
$D_{\rm T}  \simeq w^2 \tau_{\rm corr}$
formed here only with the vertical turbulence intensity.

More structure results if the turbulence is subject to a basic
rotation. The expressions  are given here only  in the
so-called $\tau$-approximation which is very close to the mixing-length
approximation (\K\ 1986). We find 
\beg
D_{ij} = D_{\rm T}\  (\Psi_1 \delta_{ij}
+ \Psi_2 \Omega_i^\circ \Omega_j^\circ),
\label{19}
\ende
with $\vec{\Omega}^\circ = \vec{\Omega}/\Omega$ as the vector
parallel to the rotation axis. The components of the tensor are 
\begin{eqnarray}
D_{rr} &=&  D_{\rm T}\  (\Psi_1 + \Psi_2 \cos^2\theta), \\
D_{\theta\theta} &=& D_{\rm T}\   (\Psi_1 + \Psi_2 \sin^2\theta),\\
D_{r\theta} &=&  D_{\rm T}\    \Psi_2
\sin\theta\cos\theta
\label{20}
\end{eqnarray}
(see Hathaway 1984). The $\Psi_2$ ensures a latitudinal transport of the 
chemical composition. 
Without rotation $\Psi_1  = 1$ and $\Psi_2 =  0$.
 The rotational quenching
 functions result as 
\beg
\Psi_1 = {3\over 4\Omega^{*2}} \bigg(1+
(\Omega^{*2}-1){\arctan\Omega^* \over \Omega^*}\bigg) 
\ende
and 
\beg
\Psi_2 ={3\over 4\Omega^{*2}} \bigg(-3 + (\Omega^{*2} + 3)
{\arctan\Omega^* \over \Omega^*}\bigg)
\label{22}
\ende
with the Coriolis number
$\Omega^* = 2 \tau_{\rm corr} \Omega$.
\subsection{Horizontal turbulence}
As usual we consider the temperature profile below the convection zone as 
stable so that any vertical fluctuations there are strongly suppressed. Any 
turbulence which possibly exists in the tachocline must be anisotropic. As a model of such 
an anisotropic turbulence, in this section a strictly horizontal turbulence is 
considered, i.e., there are, for zero rotation, no vertical 
fluctuations.

For horizontal turbulence without rotation  the  diffusion
tensor becomes  
\beg
D_{ij} = {1\over 2}  \int\!\!\!\int\limits_{\!\!\!0}^{\!\!\!\infty} {Dk^2 \over \omega^2 +
D^2 k^4} \ E_2(k,\omega) \ dk d\omega\ (\delta_{ij} - g_i
g_j)
\label{23.1}
\ende
 without any vertical components. But with rotation and within the
$\tau$-approximation we find 
\begin{eqnarray}
\lefteqn{D_{ij} =  s D_{\rm T}\  \bigg\{ \delta_{ij}
\bigg(\Psi_3 + 3(\vec{g} \cdot \vec{\Omega}^\circ)^2 \Psi_2\bigg) 
 + 3\Psi_2 \Omega_i^\circ \Omega_j^\circ + 
}\nonumber\\
&& + \Psi_4 g_i g_j + (\vec{g} \cdot \vec{\Omega}^\circ) \Psi_5 (g_i
\Omega_j^\circ + g_j \Omega_i^\circ)\bigg\}, 
\label{24}
\end{eqnarray}
which yields in spherical coordinates the tensor components 
\begin{eqnarray}
\lefteqn{D_{rr} = s D_{\rm T}\  \left(\Psi_3 + \Psi_4 +
\cos^2\theta (6\Psi_2 + 2\Psi_5)\right) , }\nonumber\\
\lefteqn{D_{\theta\theta} =    s D_{\rm T}\  (\Psi_3 +
3\Psi_2),}\nonumber\\
\lefteqn{D_{r\theta} = D_{\theta r} = -  s D_{\rm T}\   
\sin\theta\cos\theta (3\Psi_2 + \Psi_5) . }
\label{26}
\end{eqnarray} 
The amplitude $s$ is introduced as the intensity of the horizontal turbulence 
in units of the intensity of the isotropic turbulence in the bulk of the 
convection zone.

\begin{figure}
\psfig {figure=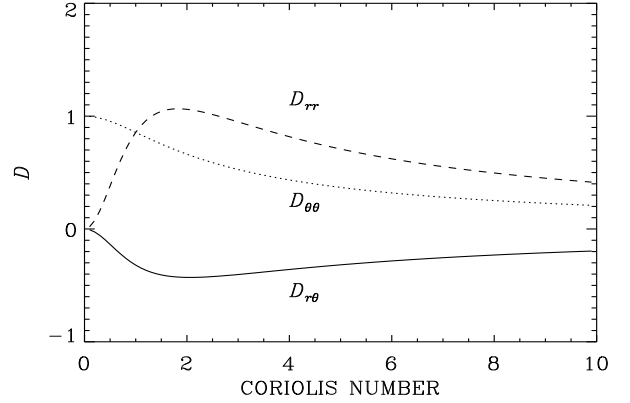,height=6truecm,width=9truecm}
\caption[]{The diffusion tensor components (\ref{26}) for midlatitudes ($\theta=45^°$) in their dependence on the Coriolis number $\Omega^*$. Note that for fast rotation ($\Omega^*\geq 1$) all components are rotationally quenched}  
\label{ss2}
\end{figure}

The rotational quenching functions result as
\begin{eqnarray}
\Psi_3 &=& {3\over 4\Omega^{*2}} \bigg(7 - (\Omega^{*2} + 7)
 {\arctan\Omega^* \over \Omega^*}\bigg) , \nonumber\\
\Psi_4 &=& {3\over 4\Omega^{*2}} \bigg(-{3\Omega^{*2} +1 \over
\Omega^{*2} +1} + (\Omega^{*2} + 1) {\arctan\Omega^*\over
\Omega^*}\bigg) , \nonumber\\  
\Psi_5 &=&  {3\over 4\Omega^{*2}} \bigg({\Omega^{*2} +3 \over
\Omega^{*2} +1} + (\Omega^{*2}-3) {\arctan\Omega^*\over
\Omega^*}\bigg) .   
\label{27.1}
\end{eqnarray}
Without rotation 
$ \Psi_3 = - \Psi_4 = 1$ and $\Psi_5 = 0$,  only the latitudinal diffusion 
coefficient $D_{\theta\theta}$ exists. With rotation, the appearance of the vertical  diffusion $D_{rr}$ and the off-diagonal components
$D_{r\theta} = D_{\theta r}$ can be observed (Fig. \ref{ss2}), which are  playing -- as we shall see
--   key roles in the 
mean-field equation for the large-scale concentration distributions in
rotating turbulence fields. One can demonstrate that this equation  for all $\Omega^*$ remains of the elliptical type.

Note that the  influence of rotation is so strong that for $\Omega^*\simeq 1$ the vertical diffusion -- which is only rotationally created -- equals the latitudinal diffusion.  Note also that the new diffusion coefficients are strongly $\theta$-dependent, resulting in $\theta$-dependent profiles of the mean concentration (see below).
\section{The model design} 
The mean-field diffusion equation without 
large-scale circulation reads 
\begin{eqnarray}
\lefteqn{\rho {\partial \bar{C} \over \partial t}
={1\over r^2}
{\partial \over \partial r}\left\{\rho r^2 \bigg(D_{rr}
{\partial \bar{C} \over \partial r} + {D_{r\theta} \over
r} {\partial \bar{C} \over \partial \theta}\bigg)\right\}+}\nonumber\\
&& + {1\over r^2 \sin\theta} {\partial \over \partial
\theta}\left\{\rho \sin\theta \bigg(D_{\theta\theta} {\partial
\bar{C} \over \partial \theta} + r D_{r\theta} {\partial
\bar{C} \over \partial r}\bigg)\right\} .
\label{28}
\end{eqnarray}
The boundary conditions are
\beg
\bar{C} = 0
\label{29}
\ende
at the lower boundary ($x_0=0.6$), where the lithium may be destroyed, and 
\beg
D_{rr} {\partial \bar{C} \over  \partial r}
+ {D_{r\theta} \over r} {\partial \bar{C} \over \partial \theta} =
0
\label{30}
\ende
at the solar surface ($x=1$) where no radial flux may be allowed.

For the correlation time in the definition of $\Omega^*$ a
radial profile is used very similar to the radial profile of
the turnover time in the mixing-length theory of the solar
convection zone, i.e.
\beg
\Omega^* = \Omega^*_{\rm i} \Bigl(\frac{x}{x_{\rm i}}\Bigr)^{7/6} 
\Bigl(\frac{1-x}{1-x_{\rm i}}\Bigr)^{3/2},
\label{32}
\ende
where  $\Omega^*_{\rm i} $ is the $\Omega^*$-value at $x_i=0.7$. Within the 
surface layer, $x>0.95$, the Coriolis number
$\Omega^*$ was put  to zero. 

Eq.~(\ref{28}) is used in  dimensionless form, distances   measured in units
of radius, $ R$, and  times  measured in units of the 
diffusion time, $\tau_{\rm diff} = R^2/D_{\rm T}$. 
\begin{figure}
\psfig{figure=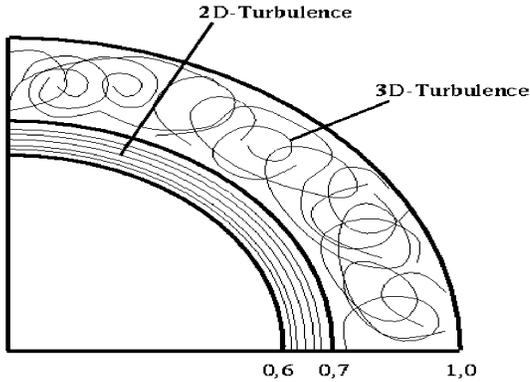,height=5truecm,width=7truecm}
\caption[]{The overall turbulence model. Below the convection
zone with isotropic turbulence there is a stable `tachocline'
layer with a horizontal turbulence field. At its bottom 
lithium is burned}  
\label{ss1}
\end{figure}
We start with a model close to that of Vincent \ea\ (1996). 
The   diffusivity  $D_{\rm T}$ may be   constant in the whole
integration region, $x_0 \leq x \leq 1$. Between the  burning
zone at $x_0$ and the bottom of the convection zone,
$x_{\rm i}$, there is a turbulence field with motions
only in the horizontal directions (Fig. \ref{ss1}). Its intensity is
given by the parameter $s$. Following Spiegel \& Zahn (1992) we
shall refer to the region $x_0 \leq x \leq x_{\rm i}$ as the solar
`tachocline'. The probing of its turbulence with respect to the lithium problem is the
main scope of the present study. 

\begin{figure}
\psfig{figure=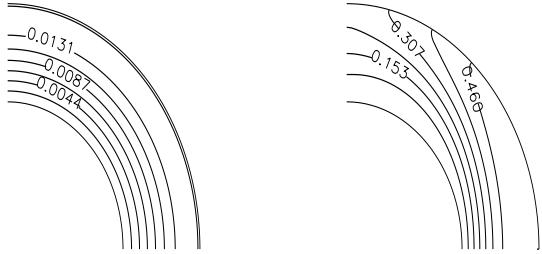,height=5truecm,width=9truecm}
\caption[]{Isolines  of the chemicals without (LEFT)
and with (RIGHT) rotation feedback on the turbulence,
$\Omega_{\rm i}^*= 6$. In both cases $s= 100$. Note that the rotation produces
  latitudinal profiles in the composition} 
\label{ss}
\end{figure}

The general influence of rotation on the spatial distribution of 
chemicals inside the star is given in Fig.  \ref{ss}  for both
$\Omega^*_{\rm i} =0$ and $\Omega^*_{\rm i} =6$. The rotational influence
produces a distinct dependence of the concentration on latitude.
\begin{figure}[tvb]
\psfig{figure=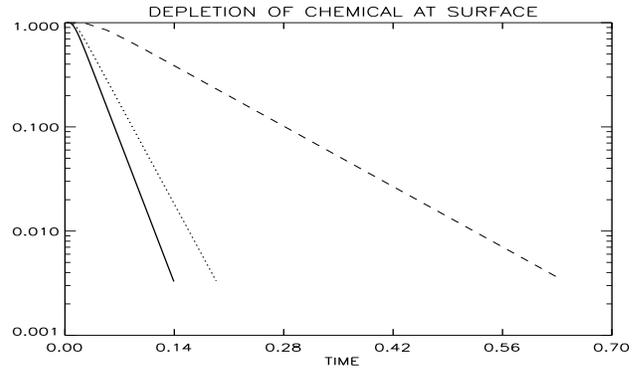,height=5truecm,width=9truecm}
\caption[]{The  decay of chemicals at the surface of the convection zone 
for  isotropic   turbulence in the whole computational region, $x_0<x<1$. 
 No rotation (solid), with rotation ($\Omega^*_{\rm i}=6$, dashed),  extra horizontal 
 turbulence in the tachocline beneath the convection zone ($\Omega^*_{\rm i}=6$, $s=100$, dotted)
}
\label{sss}
\end{figure}
Figure \ref{sss}  shows the decay  of chemicals at the
surface. The solid line  corresponds to  non-rotating
turbulence for any value of $s$. The same result is obtained even
for $s=0$, \ie without horizontal turbulence.
The result of  Michaud \& Charbonneau 
(1991) and Charbonneau (1992) is confirmed in that `horizontal turbulence cannot reduce the effect of
a given vertical turbulent diffusion coefficient'. 

The dashed curve describes the depletion of the chemicals under
the action of isotropic but rotating turbulence. Note the
rotational slow-down of the chemical depletion. 
{\em Faster} depletion, however,   is provided  by the extra
influence of rotating anisotropic  turbulence beneath the
convection zone. By this result the  influence of the  rotation
on the  turbulence field is reflected. The basic phenomenon is
the formation of a latitudinal profile of
the concentration as shown in Fig. \ref{ss}. 

We have to take from Fig. \ref{sss} that the additional anisotropy due to the
basic rotation acts in an unexpected way: rotation suppresses
the mixing  of isotropic turbulence but, on the other hand, it
enables a horizontal turbulence (which without rotation would not be
active) to accelerate the vertical mixing. The rotation in the mixing process 
which is demonstrated in Fig. \ref{sss} has a very complex character: it both 
suppresses and deforms the turbulence. The interplay between both procedures 
yields the resulting effect. We shall see that for slow rotation enhanced diffusion 
dominates while for fast rotation the diffusion reduces the suppression of 
    turbulence. As a result for a given rotation rate a maximal diffusion rate exists 
between slow and fast rotation.

\section{Lithium depletion and the dynamo problem}
Now the turbulence in the convection zone is considered as isotropic
turbulence under the influence of 
rotation described in Section 2.1. Below the convection zone
the turbulence  may be so anisotropic
that  vertical motions do not   exist. The basic rotation, however,
produces off-diagonal components in the diffusivity tensor as described in Section 
2.2.
Thus we have 2 free parameters for the
tachocline turbulence: intensity $s$ of the
horizontal motion and correlation time $\tau_{\rm corr}^{(2)}$ of the
eddies. For $s\to 0$ the turbulence
completely disappears and for   $\tau_{\rm corr}^{(2)}\to 0 $ the
rotational influence disappears. In both limits the decay time of
the chemicals at the solar surface {\it must} become infinite. 

\begin{figure}[]
\psfig{figure=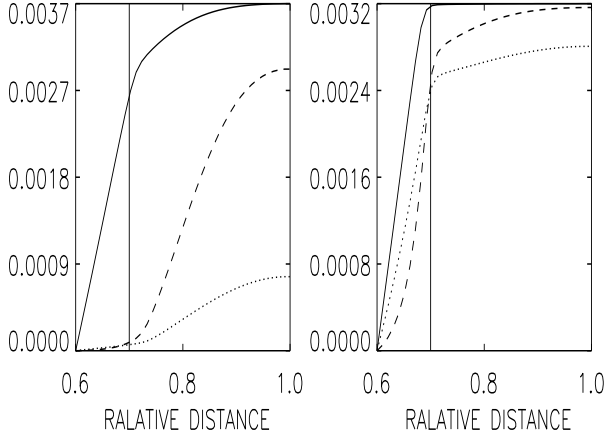,height=6truecm,width=8truecm}
\caption[]{Radial profiles of the chemical concentration along the equator (solid), mid-latitudes (dashed) and the poles (dotted) for $s=1$ (LEFT) 
and $s=0.01$ (RIGHT). Only for small horizontal intensities does the
pole-equator difference become rather small. The 
time-dependence is now very weak, $\Omega^*_{\rm i}=6$ at the base of the
convection zone and also within  the tachocline} 
\label{lat}
\end{figure}

More information is given in Fig. \ref{lat}, where the
radial dependencies of the concentration profiles inside
the star are given for three different latitudes and two
different horizontal intensities. 
The diffusion is now much slower than with isotropic turbulence in the whole 
computational, space as in Fig. \ref{sss}.
The 
differential concentrations are produced  in the tachocline and
are directly imprinted at the surface and are therefore {\em
not} screened by the convective pattern within the convection zone. This is the 
same situation as for differential rotation but different from turbulent heat transport (cf. Stix 1981). The 
time-dependence of the surface differences of the concentration
is so  weak  that  our  snapshots are always characteristic
the situation. 
\begin{figure}[]
\psfig{figure=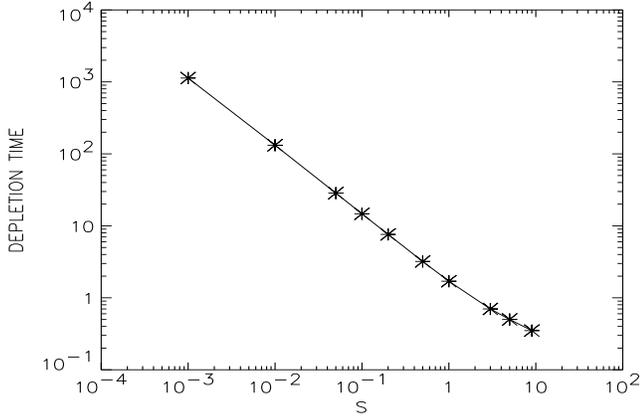,height=6truecm,width=9truecm}
\caption{
Depletion time in units of the diffusion time. For weak
horizontal  turbulence intensities it approximately varies with $1/s$.
$\Omega^*_{\rm i}=6$  at the base of the convection zone and also
within  the tachocline 
}
\label{decay}
\end{figure}
In Fig. \ref{decay} the decay time of an initially uniform
concentration is plotted for various horizontal turbulence
intensities in the tachocline. For $s=0$, of course, there is no depletion, 
but the depletion time is strikingly short for nearly homogeneous turbulence fields with (say)
$s=1$. The correlation time  profile  and the basic rotation
are taken from a solar model by Stix \& Skaley with $\Omega^*=6$ at the base of the
convection zone. We find that {\em rotation is highly effective} in
transporting the chemicals to the burning zone at $x_0$. A scaling such as
\beg
\tau_{\rm dec} \simeq \tau_{\rm diff}/s
\label{tau}
\ende
is found  compatible with the idea that the {\em diffusion of horizontal 
turbulence under the influenc of rotation with $\Omega^*\simeq 1$ can 
 be considered approximately as diffusion of isotropic turbulence}. In reality, both the characteristic times differ by 7 orders of 
magnitude. Therefore, the
horizontal rms velocity of the tachocline turbulence must  not
exceed $10^{-(3\dots 4)}$ of the convection zone turbulence.  That there
is still lithium  at the solar surface is only compatible
here with the given concept   of rotating turbulence with very small
$s$ of the tachocline turbulence. Consequently, the horizontal
turbulence intensities must be very low, i.e.
of the order of 1 cm/s. There is certainly no
chance of maintaining magnetic fields of the order of kGauss by such a
slow flow.

\begin{figure}[]
\psfig{figure=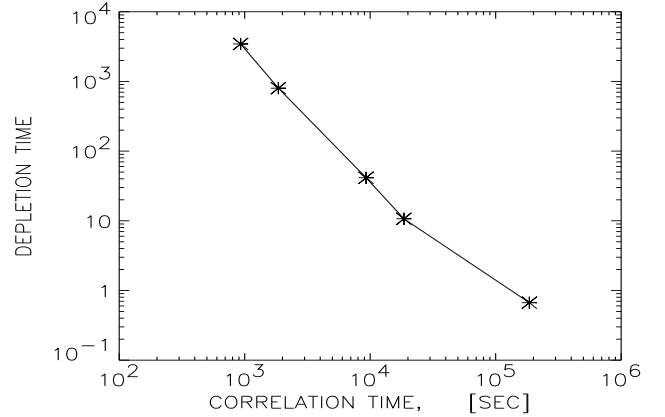,height=6truecm,width=9truecm}
\caption{
Depletion time $\tau_{\rm dec}$ (in units of the diffusion time) for  horizontal  turbulence
intensities with $s=1$ strongly depends on  the correlation
time of the  tachocline turbulence, taken here  
for a rotation period of  25 days
}
\label{taucorr}
\end{figure}

But there is also the  possibility that  the tachocline
turbulence  has a rather short correlation time, e.g.  similar to
that of
granulation at the top of the convection zone. Then the
Coriolis number, $\Omega^*$, is smaller than unity and the
rotational influence can  only be very small. For a given intensity
($s=1$) the resulting depletion times are shown in
Fig. \ref{taucorr}. We find a relation  
$\tau_{\rm dec}/ \tau_{\rm diff} \simeq 10^9/\tau_{\rm corr}^2$
with $\tau_{\rm corr}$ in seconds. Similarly, for $s=0.01$  we find   
\beg
\tau_{\rm dec} \simeq {10^{14} \over \tau_{\rm
corr}^2} \tau_{\rm diff}
\label{tt}
\ende
from our simulation.
In the latter case with $\tau_{\rm corr}\simeq 10^{3\dots 4}$ s the
lithium decay time will approach the order of $10^9$ yrs. The corresponding Coriolis number is always smaller than 0.05.
Again,  with such a small value there is no hope of forming an effective 
\alf-effect for an appropriate dynamo process. In this approach, we find  a 
close relation between the lithium problem and the theory of the solar dynamo. 
Within this framework, the observations do {\em not} favour the existence of an overshoot dynamo below the bottom of the convection zone.

\section{Rotation rate variations}
\begin{figure}[]
\psfig{figure=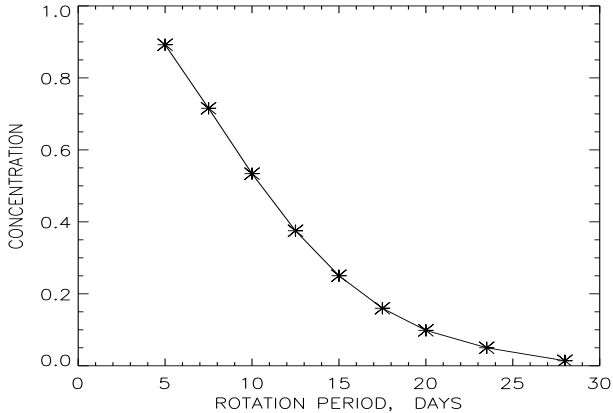,height=6truecm,width=9truecm}
\caption[]{The general situation for young stellar clusters: The same turbulence model with $s=0.01$ but the angular
velocity varied.  The surface concentration at the equator
is shown after 100 diffusion times. Due to the rotational quenching of the 
eddy diffusivity, the fast rotators possess the
highest  surface  concentration of chemicals
}
\label{cluster}
\end{figure}

For a given stellar model the basic rotation also forms a
free parameter which can be varied.  
One can vary the basic rotation rate such as it
varies in a young stellar cluster for all stars with the same
age. The question is how the lithium varies.  
The same turbulence model can be
used  within the sample. We take $s=0.01$  and
$\tau_{\rm corr}^{(2)}$ =  12 days.  Figure \ref{cluster}
shows the results. Due to the rotational quenching of the eddy diffusivity
 (see Boubnov \& Golitsin 1995), the faster the rotation the more lithium
remains  at the surface. The rotational quenching of the eddy
diffusivity dominates the effect  of the large latitudinal
gradients of the concentration. The slower rotators are thus more
effective in mixing the chemicals downwards. The plot,
however, does not display that for still slower rotation, in the
limit for $\Omega^*<1$, 
 the mixing again becomes slower and slower and therefore the decay
of the primordial lithium, too. The observation of the
lithium-rotation correlation in young stellar clusters seems to
confirm our prediction. 
\section{Conclusions}

Rotation easily produces off-diagonal elements and other new terms in the diffusion tensor. 
We have shown with our computations that under the rotational influence, even a
 {\em horizontal} turbulence field will be very  effective in transporting chemical 
 components in the vertical direction. Two main consequences of this finding have 
 been formulated, i.e.
 \begin{itemize}
 \item even a strictly horizontal turbulence in the solar tachocline must be very weak (or
 very short-lived) to ensure the observed slow diffusion of lithium downward to
 the 2.6 Mio K layer where it burns,
 \item for G-stars in young stellar clusters a correlation should exist 
 between the lithium abundance and the actual rotation rate. The faster the
 rotation, the stronger the eddy diffusivity is quenched and the higher the
 lithium concentration in the convection zone must be.
 \end{itemize}
 Our concept also provides statements for the long-term evolution
 of the lithium abundance under the influence of the general spin-down of
 solar-type stars. To this end a time-dependent code (which is in preparation) must simultaneously include
 both the time and depth-dependent effects of lithium burning and the decay law
 of the rotation rate with time.

\acknowledgements
{  V. V. Pipin acknowledges the kind support by the Deutsche
Forschungsgemeinschaft (436 RUS 113/255). The referee, J.P. Zahn, is acknowledged for many valuable suggestions concerning the concept of the paper.} 
 
\begin {thebibliography} {} 
\item{}
Ahrens B., Stix M., Thorn M., 1992, \aa 264, 673
\item{}
Baglin A., Morel P.J., Schatzman, E., 1985, \aa 149, 309
\bib
Boubnov B. M., Golitsin G. S., 1995, Convection in Rotating Fluids,
Kluwer Academic Publishers, Dordrecht, London 
\item{}
Chaboyer B., Zahn J.-P., 1992, \aa 253, 173
\item{}
Charbonneau P., Michaud G., 1991, \apj 370, 693
\item{}
Charbonneau P., 1992, \aa 259, 134
\item{}
Charbonnel C., Vauclair S., Zahn J.-P., 1992, \aa 255, 191
\item{}
Dolginov A.Z., Silantev N.A., 1992, GAFD 63, 139
\item{}
Duncan D.K., 1981, \apj 248, 651
\item{}
Hathaway D.H., 1984, \apj 276, 316
\item{}
K\"ammerer S., 2000, PhD Thesis, Potsdam 
\item{}
Kitchatinov L.L., 1986, GAFD 35, 93 
\item{}
Lebreton Y., Maeder A., 1987, \aa 175, 99
\item{}
Michaud G.J., Charbonneau P., 1991, Space Science Reviews 57, 1
\item{} R\"udiger G., 1989, Differential Rotation and Stellar Convection, 
Gordon \& Breach Sc. Pub., New York
\item{}
Schatzman E., Maeder A., 1981, \aa 96, 1
\item{}
Skumanich A., 1972, \apj 171, 565
\item{}
Spiegel E.A., Zahn J.-P., 1992, \aa 265, 106
\item{}
Spruit H.C., 1984, Mixing in the solar interior. In: Proc. of
the Fourth European Meeting on Solar Physics, The Hydrodynamics
of the Sun. Noordwijk, p. 21
\item{}
Stix M., 1981, \aa 93, 339
\item{}
Stix M., Skaley D., 1990, \aa 232, 234
\item{}
Vincent A., Michaud G., Meneguzzi M., 1996, Phys.
Fluids 8, 1312
\item{}
Zahn J.-P., 1989, Theory of transport processes. In: IAU Coll. 121, Berthomieu G. 
\& Cribier M. (eds.) Inside the Sun. Kluwer, p. 425 
\item{}
Zahn J.-P., 1992, \aa 265, 115
\end{thebibliography}{}
\end{document}